\documentclass[aps,prd,preprint,showpacs,nofootinbib,superscriptaddress]{revtex4}
\usepackage{latexsym,bm,amsmath,amssymb,xfrac,xcolor}
\DeclareMathAlphabet{\mathpzc}{OT1}{pzc}{m}{it}

\newcommand{\cD}{\mathcal{D}}
\newcommand{\cG}{\mathcal{G}}
\newcommand{\nn}{\nonumber}

\begin{document}
	
\title{Bachian Gravity in Three Dimensions}

\author{G{\"o}khan Alka\c{c}}
\email{galkac@metu.edu.tr}
\affiliation{Physics Group, Middle East Technical University,\\
	Northern Cyprus Campus, Kalkanli, via Mersin 10 Turkey}
\affiliation{Department of Physics, Faculty of Arts and  Sciences,\\
	Middle East Technical University, 06800, Ankara, Turkey}

\author{Mustafa Tek}
\email{mustafa.tek@medeniyet.edu.tr}
\affiliation{Department of Physics, Faculty of Arts and  Sciences,\\
	Middle East Technical University, 06800, Ankara, Turkey}
\affiliation{Istanbul Medeniyet University, Faculty of Engineering and Natural Science,\\
	 Engineering Physics Department, TR-34730 Istanbul, Turkey}

\author{Bayram Tekin}
\email{btekin@metu.edu.tr}
\affiliation{Department of Physics, Faculty of Arts and  Sciences,\\
	Middle East Technical University, 06800, Ankara, Turkey}

\date{\today}

\begin{abstract}
In three dimensions, there exist  modifications of Einstein's gravity akin to the topologically massive gravity that describe massive gravitons about maximally symmetric backgrounds. These theories are built on the three-dimensional version of the Bach tensor (a curl of the Cotton-York tensor) and its higher derivative generalizations; and they are on-shell consistent without a Lagrangian description based on the metric tensor alone. We give a generic construction of these models, find the spectra and compute the conserved quantities for the Banados-Teitelboim-Zanelli black hole. 
\end{abstract}
\maketitle
\section{INTRODUCTION}

It would be pedantic to stress the importance of Einstein metrics (Ric=$\lambda g$): in four dimensions, to the best of the present day knowledge, the Universe without matter is locally an Einstein manifold with all the interesting stuff (such as black holes, their mergers and gravitational waves). However, even after more than a century's work, we still do not have a good grip of the Einstein metrics in four dimensions and beyond. This state of affairs affects our understanding of some problems of classical gravity; but, more importantly it complicates a possible construction of the quantum version of the theory. For this purpose, the ($2+1$)-dimensional gravity, which is locally much simpler, has always attracted attention. But it is easy to see that pure general relativity (GR) in $2+1$ dimensions is locally too simple to be of much help: locally Einstein metrics are Riemann flat (or constant curvature) since in this dimension we have the following identity
\begin{equation}
R_{\mu\alpha\nu\beta}=\epsilon_{\mu\alpha\sigma}\epsilon_{\nu\beta\rho}G^{\sigma\rho},
\end{equation}
where $\epsilon_{\mu\alpha\sigma}$ is totally antisymmetric tensor and $G_{\rho\sigma}$ is the Einstein tensor $G_{\rho\sigma}=R_{\rho\sigma}-\frac{1}{2}g_{\rho\sigma}R$. This basically says that in a vacuum there is no gravity, and no gravitation. When a negative cosmological constant is introduced, local triviality is not lifted, but there is the all important 
Ba\~{n}ados-Teitelboim-Zanelli (BTZ)  black hole \cite{btz} that can carry mass, spin and pretty much all the properties of its four-dimensional analog Kerr black hole, save the curvature singularity and the speed-of-light surface. So some of the Einstein metrics are highly nontrivial (when considered in $2+1$ GR) but one of course  still needs local nontriviality, gravitation, gravitational waves {\it etc.} to be able to learn something from this lower-dimensional setting. 

Fortunately, this can still be achieved with Einstein metrics but not as solutions to GR but as solutions to modified gravity theories, such as the topologically massive gravity (TMG) \cite{djt}, new massive gravity (NMG) \cite{nmg,morenmg} or Born-Infeld extension of NMG \cite{binmg}. All these theories accommodate Einstein metrics and more general metrics that are not Einstein. But the good thing is that in these theories, perturbation about an Einstein metric can be interpreted as gravitons (usually massive) or gravitational waves. Hence these theories are much richer than Einstein's pure $2+1$ GR and simpler than the $3+1$ GR. The immediate aim is to be able to define and understand a version of quantum gravity in a $2+1$-dimensional setting. For this purpose, our current best hope is the AdS/CFT duality \cite{Maldacena} which reduces the problem to a construction of a two-dimensional boundary conformal field theory for the AdS bulk of a given $3D$ theory.

In this context, what we currently know can be summarized as follows: NMG (a nonlinear extension of the Fierz-Pauli massive spin-2 model) does not provide such a theory: it is unitary either in the bulk or on the boundary \cite{morenmg} and so suffers from the so-called ``bulk-boundary unitarity clash".  In fact it was proven in \cite{GST_all_unitary} that no theory that has the same particle content as NMG can be bulk and boundary unitary at the same time. This is a strong theorem which also rules out any $f(\text{Ricci})$-type higher curvature extensions of the NMG such as the cubic and quartic theories  obtained by demanding the existence of a holographic $c$-function in \cite{sinha,paulos} and the infinite order Born-Infeld extension \cite{binmg}. On the other hand, TMG is different, it falls out of this ``no go'' theorem as it has a ``single" massive spin-2 graviton (with either positive or negative helicity). But we know that except for the ``chiral" point, where the topological mass ($\mu$) and AdS radius ($\ell$) are related as $\mu \ell=1$, this theory cannot be unitary in the bulk and on the boundary \cite{strom,strom2}. That leaves us with the chiral gravity case only which needs a longer discussion; but, let us just note that at the chiral point at first sight the theory seems to be bulk and boundary unitary but then exactly at those parameter values of the theory, the linearized equations has a ghost like new mode \cite{grumiller}. This new mode can be dual to an operator in a log-CFT which is non-unitary. So, in trying to get a viable dynamical theory of $2+1$ dimensional gravity, we seem to be hovering in limbo. 
But it was argued in \cite{strom2} and \cite{carlip} that the log mode may not survive linearization; namely, it is an artifact of the linearized theory and does not come from the linearization of an exact solution. In fact this expectation was proven to be true recently \cite{altas1,altas2}. Therefore, the status of the chiral gravity now is that it is a potentially viable classical and quantum theory; but, one must still show the latter by actually finding the corresponding CFT on the boundary.

To overcome the bulk-boundary unitarity clash of the 3D theories, an interesting idea was put forward in \cite{townsend} where the authors introduced the so-called minimal massive gravity (MMG). The crux of the idea is that instead of a Lagrangian, based on the metric only, one can define the theory with the field equations that are on-shell consistent (see also \cite{townsend3} for a discussion of the main idea). A detailed analysis of the MMG theory \cite{Gokhan_mmg} showed that, just like TMG, the theory is free of the bulk-boundary unitarity clash only at the chiral point\footnote{There is an important caveat here: in \cite{Gokhan_mmg}, the unitarity analysis of MMG was done in the linearized theory and in the metric formulation, where there is no non-linear action. In the first order formulation, where there is a non-linear action, MMG seems to be free of the bulk-boundary unitarity clash \cite{townsend}.} \cite{Tekin_chiral,Alishahiha_chiral}. The matter coupling in such theories was achieved in \cite{townsend2} and another on-shell consistent theory named exotic massive gravity (EMG) was recently given in \cite{townsend4}. Some solutions of this theory were given in \cite{Giribet-sol}. Such on-shell consistent theories offer interesting possibilities: a cursory look may lead one to think that these theories are too unwieldy, but this is not the case as we shall explore some further such theories here.   

The layout of the paper is as follows. In the next section, we give a construction of the 3D Bachian gravity. In section III, we consider the version of the theory coming from quadratic gravity, in Section IV we construct the conserved charges and compute them for the rotating BTZ metric. 

\section{3D Bach Tensor and On-shell Consistency}

Let us go back to the discussion of Einstein metrics that was alluded to above: perhaps the next ``nice" set of metrics are the ones conformally related to the Einstein metrics. Succinctly stated the problem is this: given a metric $g$ (which is not necessarily Einstein) can one construct a metric, $\tilde{g}\equiv\Omega^2 g$, which is Einstein given that $\Omega$ is smooth and $\Omega>0 $? In $n-$dimensions, the generic necessary and sufficient conditions for such a metric $\tilde{g}$ to exist are too difficult to handle. But, in four dimensions the problem simplifies a little bit in the sense that the necessary condition is the vanishing of the so-called ``Bach Tensor"
\begin{equation}
B_{\mu\nu}\equiv \big(\nabla^{\alpha}\nabla^{\beta}+\frac{1}{2}R^{\alpha\beta} \big)W_{\mu\alpha\nu\beta},
\label{bach_tensor}
\end{equation}
where $W_{\mu\alpha\nu\beta}$ is the Weyl tensor. The Bach tensor is symmetric, traceless $B \equiv g^{\mu\nu}B_{\mu\nu}=0$, divergence-free $\nabla^{\mu}B_{\mu\nu}=0$ and conformally invariant (in four dimensions). Moreover, one can show that $B_{\mu\nu}$ comes from the variation of the action
\begin{equation}
S=\int d^4 x \sqrt{-g} W_{\mu\nu\alpha\beta}W^{\mu\nu\alpha\beta}.
\end{equation}
This so-called conformal gravity admits all the Einstein metrics as solutions, but there are non-Einstein solutions. Remarkably, with some simple (Neumann) boundary conditions, one can show that out of all the Bach flat manifolds, only
Einstein manifolds can be selected \cite{malda_conformal} . 

One can naturally wonder the simpler problem, that is, the problem of the conformal Einstein metrics in three-dimensions. As the Weyl tensor vanishes identically in three-dimensions, the naive dimensional continuation of the Bach tensor as defined by (\ref{bach_tensor}) to three dimensions does not yield any further information. But as was realized in \cite{Tekin_mmg2,altas3}, using the 3-index Cotton tensor as a potential to the Weyl tensor yields a meaningful 3D Bach tensor. Recall that the $n$-dimensional Cotton tensor is 
\begin{equation}
C_{\alpha\mu\nu}=\nabla_{\alpha}R_{\mu\nu}-\nabla_{\mu}R_{\alpha\nu}-\frac{1}{2(n-1)}\big( g_{\mu\nu}\nabla_{\alpha}R-g_{\alpha\nu}\nabla_{\mu}R \big),
\end{equation}
which is antisymmetric in the first two indices.  This tensor is conformally invariant only in three dimensions. Using this, we define the analog of the $n$-dimensional Bach tensor as
\begin{equation}
B_{\mu\nu}\equiv \frac{1}{2}\nabla^{\alpha}C_{\alpha\mu\nu}+\frac{1}{2}R_{\alpha\beta}W_{\mu}{^{\alpha}}{_{\nu}}{^{\beta}}.
\end{equation}
In particular, for $n=3$, we can express the Cotton tensor in terms of the Cotton-York tensor ($C_{\mu\nu} \equiv \epsilon_{\mu}\,^{\sigma \rho} \nabla _\sigma S_{\rho\nu}$ with $S_{\mu \nu} = R_{\mu \nu} - \frac{1}{4} g_{\mu \nu} R$.) as
\begin{equation}
C^{\sigma\rho}{_{\nu}}=-\epsilon^{\sigma\rho\mu}C_{\mu\nu}
\end{equation}
where
\begin{equation}
C_{\mu\nu}\equiv\frac{1}{2}\epsilon_{\mu}{^{\alpha\beta}}C_{\alpha\beta\nu}.
\end{equation}
Therefore, the 3D Bach tensor can be defined as~\footnote{To conform with the original definition \cite{Tekin_mmg2} where the tensor was denoted as $H_{\mu \nu}$, we drop an overall factor of 1/2.}  
\begin{equation}
B_{\mu\nu}\equiv \frac{1}{2}\epsilon_{\mu}{^{\alpha\beta}}\nabla_{\alpha}C_{\beta\nu}+\frac{1}{2}\epsilon_{\nu}{^{\alpha\beta}}\nabla_{\alpha}C_{\beta\mu}.
\label{3Dbach}
\end{equation}
The Cotton-York tensor plays the role of the Weyl tensor in 3D: namely it vanishes if and only if the metric is conformally flat. But an interesting situation arises in 3D: unlike the Weyl tensor (a four-index object) that does not come from the variation of an action, the Cotton-York tensor does come from the variation of the topological Chern-Simons action and it behaves regularly: $\tilde{C}^{\mu \nu}( \tilde {g}) = \Omega^{-2} C^{\mu \nu}({g})$ under conformal transformations. This says that conformally flat metrics in 3D are conformally Einstein. So, the 3D Bach tensor vanishes for conformally Einstein metrics. It is possible that its vanishing can be a sufficient condition, which we do not know. What is interesting is that, even though $B_{\mu\nu}$ (\ref{3Dbach}) is  symmetric and traceless ($B\equiv g^{\mu\nu}B_{\mu\nu}=0$), it is not divergence-free. In fact one has 
\begin{equation}
\nabla_{\mu}B^{\mu\nu}=\epsilon^{\nu\alpha\beta}R_{\alpha\sigma}C_{\beta}{^{\sigma}},
\label{divergence_bach}
\end{equation}
which vanishes for Einstein metrics and/or conformally flat or Einstein metrics. This also says that, the 3D Bach tensor cannot come from the variation of an action. In fact, one has the following variational result \cite{Tekin_mmg2}
\begin{equation}
\delta \int d^3 x \sqrt{-g} \left ( R_{\mu \nu} R^{\mu \nu} - \frac{3}{8} R^2 \right) = \int d^3 x \sqrt{-g} \left(J_{\mu \nu}+ B_{\mu \nu}\right) \delta g^{\mu\nu},
\label{K-action}
\end{equation}
with
\begin{equation}
 J_{\mu \nu} =\frac{1}{2}\epsilon{_{\mu}}^{\alpha\beta}\epsilon{_{\nu}}^{\rho\sigma}S_{\alpha\rho}S_{\beta\sigma}.
\end{equation}
One has $\nabla_{\mu}B^{\mu\nu} = -\nabla_{\mu}J^{\mu\nu}$ and $J \equiv g^{\mu \nu} J_{\mu \nu} =  R_{\mu \nu} R^{\mu \nu} - \frac{3}{8} R^2 $. So the variation of the purely quadratic theory with the NMG coefficients (this is the $K$ theory introduced in \cite{Deser_K}) naturally splits into two parts: the Bach tensor and the $J$ tensor; and the latter does not have the derivatives of the curvature. With this rather natural splitting in hand, one can deform Einstein's theory or TMG with these new tensors $J_{\mu \nu}$ and $B_{\mu \nu}$ which have been done to obtain MMG and MMG$_2$ as on-shell consistent theories. Now our task is to extend these models. 

First, let us now find some generalizations of the 3D Bach tensor (\ref{3Dbach}) and use them to construct on-shell conserved theories. Consider a 2-tensor ${\cal{E}}_{\mu\nu}$ that comes from the variation of an action such that $\nabla^{\mu}{\cal{E}}_{\mu\nu}=0$ and assume that we have a symmetric 2-tensor $\Phi_{\mu\nu}$ that does not come from the variation of an action and  $\nabla_\mu \Phi^{\mu\nu} \ne 0$.
Now, consider the following potentially viable on-shell consistent equations
\begin{equation}
{\cal{E}}_{\mu\nu}+\frac{1}{\mu}\epsilon_{\mu}{^{\alpha\beta}}\nabla_{\alpha}\Phi_{\beta\nu}+\frac{k}{\mu^2}\epsilon_{\mu}{^{\alpha\beta}}\epsilon_{\nu}{^{\sigma\rho}}\Phi_{\alpha\sigma}\Phi_{\beta\rho}=0,
\end{equation}
where $\mu$ and $k$ are parameters at this stage, but $k$ will be fixed from consistency. Inspired by the construction of MMG, this form of the field equations was first introduced in \cite{townsend4}, where the authors choose $\Phi_{\mu\nu}=C_{\mu\nu}$ to obtain EMG. The middle term is a generalization of the Bach tensor, while the last term is a generalization of the $J$ tensor.
The first and the third terms are symmetric under the interchange of indices $\mu$ and $\nu$. The second one is symmetric only if 
\begin{equation}
 \nabla_{\sigma}\Phi=\nabla_{\alpha}\Phi_{\sigma}{^{\alpha}},
\end{equation}
where $\Phi \equiv g^{\mu \nu}\Phi_{\mu \nu}$. This is the first condition on the theory. Another condition comes from the vanishing of the divergence which yields 
\begin{equation}
\nabla_{\nu}\bigg( {\cal{E}}^{\mu\nu}+\frac{1}{\mu}\epsilon^{\mu\alpha\beta}\nabla_{\alpha}\Phi_{\beta\nu}+\frac{k}{\mu^2}\epsilon^{\mu\alpha\beta}\epsilon^{\nu\sigma\rho}\Phi_{\alpha\sigma}\Phi_{\beta\rho} \bigg)=\frac{1}{\mu}\epsilon^{\mu\alpha\beta}\Phi_{\beta\lambda}\left(R_{\alpha}{^{\lambda}}+\frac{2k}{\mu}\epsilon_{\alpha}\,^{\beta \gamma}\nabla_{\beta}\Phi_{\gamma}\,^\lambda\right).
\end{equation}
Clearly this expression is not generically zero and the theory is generically inconsistent. But the explicit expression tells as that we must include Einstein's gravity in the ${\cal{E}}_{\mu\nu}$ in order to have any hope of constructing an on-shell-consistent theory; hence, we choose
\begin{equation}
{\cal{E}}_{\mu\nu} \equiv R_{\mu\nu}-\frac{1}{2}g_{\mu\nu}R+\Lambda_0 g_{\mu\nu},
\end{equation}
and then the following equation:
\begin{equation}
R_{\mu\nu}-\frac{1}{2}g_{\mu\nu}R+\Lambda_{0}g_{\mu\nu}+\frac{1}{\mu}\epsilon_{\mu}{^{\alpha\beta}}\nabla_{\alpha}\Phi_{\beta\nu}+\frac{1}{2\mu^2}\epsilon_{\mu}{^{\alpha\beta}}\epsilon_{\nu}{^{\sigma\rho}}\Phi_{\alpha\sigma}\Phi_{\beta\rho}=0\label{eqphi}
\end{equation}
with any $\Phi_{\mu\nu}=\Phi_{\nu\mu}$  satisfying $\nabla_{\mu}\Phi^{\mu}{_{\nu}}=\nabla_{\nu}\Phi$, is consistent. Observe that consistency required the constant $ k= 1/2$.  

The next obvious question is how to find a 2-tensor  $\Phi_{\mu}{_{\nu}}$  that satisfies the desired properties. This is also remarkably simple to answer: consider any action, vary it with respect to the metric and obtain a 2-tensor which is covariantly conserved. Let us call this tensor to be $\Psi_{\mu \nu}$, and then one can choose \cite{townsend4}
\begin{equation}
\Phi_{\mu\nu}:=\Psi_{\mu\nu}-\frac{1}{2}g_{\mu\nu}\Psi \quad , \quad \Psi=g^{\mu\nu}\Psi_{\mu\nu},
\end{equation} 
which satisfies the desired property  $\nabla_{\sigma}\Phi=\nabla_{\alpha}\Phi_{\sigma}{^{\alpha}}$.
Using the $\Psi_{\mu\nu}$ field, we can recast (\ref{eqphi}) as 
\begin{equation}
R_{\mu\nu}-\frac{1}{2}g_{\mu\nu}R+\Lambda_{0}g_{\mu\nu}+\frac{1}{\mu}\epsilon_{\mu}{^{\alpha\beta}}\nabla_{\alpha}\big( \Psi_{\beta\nu}-\frac{1}{2}g_{\beta\nu}\Psi \big)+\frac{1}{2\mu^2}\bigg( g_{\mu\nu}\big( \Psi_{\alpha\beta}^{2}-\frac{3}{4}\Psi^2 \big)+\Psi_{\mu\nu}\Psi-2\Psi_{\mu\alpha}\Psi_{\nu}{^{\alpha}} \bigg)=0.\label{eqpsi}
\end{equation} 
So the upshot is that we can deform Einstein's gravity  with any covariantly conserved 
$\Psi_{\mu\nu}$ in such a way that we get a nontrivial on-shell-consistent theory. 

One might wonder if one can further deform (\ref{eqphi}) or (\ref{eqpsi}) with ${\cal{O}}(\Phi^3)$ and ${\cal{O}}(\Phi^4)$ terms. Even though we have not done this for this general case, for the MMG case, where $\Phi_{\mu \nu}= S_{\mu \nu}$, it was shown in \cite{altas3} that no further terms can be added. On-shell consistency is highly restrictive and truncates the theory at the second order.

\section{Quadratic Gravity}
So far, we have only proved the consistency of field equations, whose final form are given in (\ref{eqpsi}). For the general construction of theories with {\it only} spin-2 modes and no extra scalar mode, we need to study the linearized equations around the AdS$_3$ spacetime. Although, the tensor $\Psi_{\mu \nu}$ can be chosen to be any tensor derived from an action for the consistency of the field equations, the absence of the scalar mode puts further restrictions. We start our analysis by considering an action with the quadratic curvature terms. As we shall see, it allows us to study wider range of possibilities where the tensor $\Psi_{\mu \nu}$ is derived from an action which is an arbitrary function of the Ricci tensor $f(\text{Ricci})$. Therefore, let us first consider the following action
\begin{equation}
S=\frac{1}{16\pi G} \int d^3 x \sqrt{-g} \left( \sigma R + \alpha R^2 + \beta R_{\mu\nu}^2  \right),\label{quadac}
\end{equation}
whose variation yields
\begin{equation}
	\delta S=\frac{1}{16\pi G} \int d^3 x \sqrt{-g}\, \Psi_{\mu\nu}\, \delta g^{\mu\nu},
\end{equation}
where
\begin{eqnarray}
\Psi_{\mu\nu}&=&\sigma G_{\mu\nu}+\alpha \bigg( 2RR_{\mu\nu}-\frac{1}{2}g_{\mu\nu}R^2+2g_{\mu\nu}\square R -2\nabla_{\mu}\nabla_{\nu}R \bigg)\\
&& +\beta \bigg( \frac{3}{2}g_{\mu\nu}R_{\rho\sigma}R^{\rho\sigma}-4R{_{\mu}}^{\rho}R_{\nu\rho}+\square R_{\mu\nu}+\frac{1}{2}g_{\mu\nu}\square R - \nabla_{\mu}\nabla_{\nu}R+3RR_{\mu\nu}-g_{\mu\nu}R^2 \bigg).\label{quadeq}\nonumber
\end{eqnarray}
Since it is derived from the variation of an action, the tensor $\Psi_{\mu\nu}$ is symmetric, covariantly conserved, and therefore yields consistent field equations. We now consider the linearization around the AdS$_3$ spacetime as
\begin{equation}
	g_{\mu\nu}=\bar{g}_{\mu\nu}+h_{\mu\nu},\label{lin}
\end{equation}
where the background AdS$_3$ metric satisfies
\begin{eqnarray}
\bar{R}_{\mu\nu\rho\sigma}=\Lambda\big( \bar{g}_{\mu\rho} \bar{g}_{\nu\sigma}-\bar{g}_{\mu\sigma}\bar{g}_{\nu\rho}  \big), \quad  \quad \bar{R}_{\mu\nu}=2\Lambda \bar{g}_{\mu\nu}, \quad  \bar{R}=6\Lambda, \quad \bar{G}_{\mu\nu}=-\Lambda \bar{g}_{\mu\nu},
\end{eqnarray}
and the tensor $h_{\mu\nu}$ describes the perturbations around the AdS$_3$ background. The linearized versions of Ricci tensor, Ricci scalar and the cosmological Einstein tensor are given, respectively, by
\begin{eqnarray}
	R_{\mu\nu}^L&=&\bar{\nabla}^{\rho}\bar{\nabla}_{(\mu} h_{\nu)\rho}-\frac{1}{2}\bar{\square}h_{\mu\nu}-\frac{1}{2}\bar{\nabla}_{\mu}\bar{\nabla}_{\nu}h,\nn\\
	R^L&=&-\bar{\square}h+\bar{\nabla}^{\rho}\bar{\nabla}^{\sigma}h_{\rho\sigma}-2\Lambda h,\nn\\
	\mathcal { G } _ { \mu \nu } &\equiv& \left( G _ { \mu \nu } + \Lambda g _ { \mu \nu } \right) ^ { L }=R_{\mu\nu}^L-\frac{1}{2}\bar{g}_{\mu\nu}R^L-2\Lambda h_{\mu\nu}.
\end{eqnarray}
Under the linearization (\ref{lin}), the background value of the tensor $\Psi_{\mu \nu}$ is given by
\begin{equation}
 \bar{\Psi}_{\mu\nu}=a \bar{g}_{\mu\nu},\quad  \quad a=-\Lambda\sigma+2\Lambda^2 \big( 3\alpha+\beta\big), \label{parameter_a}
\end{equation}
and its linearization yields
\begin{eqnarray}
\Psi_{\mu\nu}^L&=&\bar{\sigma}\mathcal{G}_{\mu\nu}+\big(2\alpha+\beta\big)\bigg(\bar{g}_{\mu\nu}\bar{\square}-\bar{\nabla}_{\mu}\bar{\nabla}_{\nu} +2\Lambda\bar{g}_{\mu\nu} \bigg)R^L\nonumber\\
&& + \beta\bigg( \bar{\square}\mathcal{G}_{\mu\nu}-\Lambda\bar{g}_{\mu\nu}R^L \bigg)+a h_{\mu\nu},\label{psiL}
\end{eqnarray}
with
\begin{equation}
\quad \bar{\sigma}=\sigma+12\Lambda\alpha+2\Lambda\beta. 
\end{equation}
We will also need the linearization of its trace $\Psi^L \equiv (g^{\mu\nu} \Psi_{\mu\nu})^L$, which can be computed as
\begin{eqnarray}
	\Psi^L=\bigg( 4\alpha+\frac{3}{2}\beta \bigg)\bar{\square}R^L+\bigg( -\frac{\sigma}{2}+2\Lambda \big( 3\alpha+\beta \big) \bigg)R^L.\label{eqpsit}
\end{eqnarray}

In the next section, we will constrain the parameters ($\sigma$, $\alpha$, $\beta$) by requiring the existence of only the spin-2 modes in the theory. Before we engage in that discussion, let us first explain the importance of the quadratic Lagrangian for obtaining a wider range of theories with this property. As shown in \cite{GST_all_unitary}, for any action which is given as an arbitrary function of the Ricci tensor $f(\text{Ricci})$, one can obtain an {\it equivalent quadratic action} which yields the same linearized equations. Once we determine the quadratic action with the desired properties, all the theories having this action as the equivalent quadratic action will have the same nice properties. For example, the cubic action
\begin{equation}
I = \int d ^ { 3 } x \sqrt { - g }  \left[\tilde{\sigma} \left( R - 2 \tilde{\lambda} _ { 0 } \right) +\tilde{\alpha}R ^ { 2 }+\tilde{\beta} R _ { \mu\nu } ^ { 2 } +a _ { 1 } R ^ { \mu }_{\nu} R ^ { \nu }_{\rho} R ^ { \rho }_{\mu} + a _ { 2 } R R _ { \mu\nu } ^ { 2 } + a _ { 3 } R ^ { 3 }\right],
\end{equation}
and the quadratic action
\begin{equation}
I = \int d ^ { 3} x \sqrt { - g } \left[ \sigma \left( R - 2  \lambda  _ { 0 } \right) +  \alpha  R ^ { 2 } + \beta R _ { a b } ^ { 2 } \right],\label{quad}
\end{equation}
yield the same linearized equations if their parameters are related by the following equations
\begin{eqnarray}
\sigma &=& \tilde{\sigma} -  12 \Lambda ^ { 2 } \left(  a _ { 1 } + 3 a _ { 2 } + 9 a _ { 3 } \right),\nn \\
\lambda  _ { 0 } &=& \frac { \tilde { \sigma } } { \sigma } \tilde{\lambda }_0 +  \Lambda  \left( 1 - \frac { \tilde { \sigma } } { \sigma } \right),\nn \\
\alpha  &=& \tilde { \alpha } + 2 \Lambda  \left( 2 a _ { 2 } + 9 a _ { 3 } \right), \nn \\
\beta  &=& \tilde { \beta } + 6\Lambda \left(a _ { 1 } +  a _ { 2 } \right).
\end{eqnarray}
Although a cosmological constant $\lambda_{ 0 }$ is introduced in the equivalent quadratic action (\ref{quad}), it yields a term proportional to the metric tensor in $\Psi_{\mu \nu}$ (\ref{quadeq}), which as a result shifts the parameter $\Lambda_{0}$ in the field equations (\ref{eqpsi}). The change in the parameter $\Lambda_{0}$ is trivial in our subsequent discussion and indeed one can obtain infinitely many higher curvature actions of $f(\text{Ricci})$ type whose variation gives a $\Psi_{\mu \nu}$ tensor leading to a pure spin-2 theory.

\section{Bachian Gravity}
In this section, we constrain the coefficients in the most generic quadratic action (\ref{quadac}) such that the field equations (\ref{eqphi}) describe spin-2 modes only. For this purpose, we consider the trace of the field equations which is given by
\begin{equation}
R-6\Lambda_0+\frac{1}{\mu^2}\bigg(\Phi^2-\Phi_{\mu\nu}\Phi^{\mu\nu}\bigg)=0,
\end{equation}
which, in terms of the $\Psi_{\mu \nu}$ tensor, reads
\begin{equation}
R-6\Lambda_0+\frac{1}{\mu^2}\bigg(\frac{1}{2}\Psi^2-\Psi_{\mu\nu}^2\bigg)=0.
\end{equation}
 Using the  equality $\bar{ g }^{\mu\nu}\Psi_{\mu\nu}^L=\Psi^L+a h$, linearization of the last equation yields
\begin{equation}
R^L+\frac{a}{\mu^2}\Psi^L=0.
\end{equation}
The expression for $\Psi^L$ was given in (\ref{eqpsit}),  making use of that one finds an wave equation for $R_L$:
\begin{equation}
R^L+\frac{a}{\mu^2} \bigg[ \bigg( 4\alpha+\frac{3}{2}\beta \bigg)\bar{\square}R^L+\bigg( -\frac{\sigma}{2}+2\Lambda \big( 3\alpha+\beta \big) \bigg)R^L  \bigg]=0.
\end{equation}
In order to avoid the propagating scalar mode, we should set the coefficient of the $\bar{\square}R^L$ term to zero, which yields two possibilities:
\begin{equation}
4\alpha+\frac{3}{2}\beta=0, \quad \text{or} \quad a=\Lambda\big( -\sigma+6\Lambda\alpha+2\Lambda\beta \big)=0.\label{sol}
\end{equation}
In both cases, we have $R^L=0$, and as a result we can choose the compatible transverse-traceless (TT) gauge ($\bar{\nabla}^\mu h_{\mu\nu}=0=h$).

Having studied the linearization of the trace equation and the constraints coming from the absence of the scalar mode,  we can now linearize the full field equations (\ref{eqphi}) to find the particle content of the theory and their masses. The background value the tensor $\Phi_{\mu\nu}$ is given as
\begin{equation}
\bar{\Phi}_{\mu\nu} = -\frac{a}{2} \bar{g}_{\mu\nu},
\end{equation}
and its linearization yields
\begin{eqnarray}
\Phi_{\mu\nu}^L=\Psi_{\mu\nu}^L-\frac{1}{2}h_{\mu\nu}\bar{\Psi}-\frac{1}{2}\bar{g}_{\mu\nu}\Psi^L.
\end{eqnarray}
The vacuum equation determining the effective cosmological constants is
\begin{equation}
\Lambda_0-\Lambda-\frac{a^2}{4 \mu^2}=0,
\end{equation}
where, of course, $a$ is given in (\ref{parameter_a}). The linearization of the field equations can be obtained as
\begin{eqnarray}
	\mathcal{G}_{\mu\nu}+\left(\Lambda_0-\Lambda+\frac{a^2}{4 \mu^2} \right)h_{\mu\nu}-\frac{a}{2 \mu^2}\Psi_{\mu\nu}^L+\frac{1}{\mu}\bar{\epsilon}{_{(\mu}}^{\alpha\beta}\bar{\nabla}_{|\alpha}\Psi_{\beta|\nu)}^L-\frac{a}{\mu}\bar{\epsilon}{_{(\mu}}^{\alpha\beta}\bar{\nabla}_{|\alpha}h_{\beta|\nu)}=0,
\end{eqnarray}
which looks like a complicated equation, but it can be handled with several observations. 
Using $\Psi_{\mu\nu}^L$ (\ref{psiL}) in the TT gauge, one has 
\begin{eqnarray}
\Psi_{\mu\nu}^L&=&\bar{\sigma}\mathcal{G}_{\mu\nu} + \beta\bigg( \bar{\square}\mathcal{G}_{\mu\nu}-\Lambda\bar{g}_{\mu\nu}R^L \bigg) +a h_{\mu\nu},
\end{eqnarray}
which reduces the field equations to a fifth-order equation in $h_{\mu \nu}$:
\begin{eqnarray}
    \big(1-\frac{\bar{\sigma} a}{2\mu^2}\big)	\cG_{\mu\nu}+\frac{\bar{\sigma}}{\mu}\bar{\epsilon}{_{\mu}}^{\alpha\beta}\bar{\nabla}_{\alpha}\cG_{\beta\nu}-\frac{\beta a}{2 \mu^2}\bar{\square}\mathcal{G}_{\mu\nu}+\frac{\beta}{\mu}\bar{\epsilon}{_{\mu}}^{\alpha\beta}\bar{\nabla}_{\alpha}\bar{\square}\cG_{\beta\nu}=0.\label{eqLin}
\end{eqnarray}
In order to identify the spin-2 modes, we introduce the mutually commuting operators \cite{strom}
\begin{eqnarray}
\big(\cD^{L/R}\big){_{\mu}}^{\nu}&:=&\delta{_{\mu}}^{\nu}\pm \ell \bar{\epsilon}{_{\mu}}^{\alpha\nu}\bar{\nabla}_{\alpha},\nn\\
\left( \mathcal { D } ^ { p _ { i } } \right) _ { \mu } ^ { \nu } &:=& \delta _ { \mu } ^ { \nu } + \frac { 1 } { p _ { i } } \bar{\epsilon}{_{\mu}}^{\alpha\nu}\bar{\nabla}_{\alpha},\qquad i = 1, 2, 3.\label{operators}
\end{eqnarray}
In the TT gauge, we have $\bar{\nabla}^{\rho}\bar{\nabla}_{\mu}h_{\rho\nu}=-\frac{3}{\ell^2}h_{\mu\nu}$ and the linearized cosmological Einstein tensor can be written as
\begin{equation}
	\cG_{\mu\nu}=-\frac{1}{2}\big(\bar{\square} +\frac{2}{\ell^2}\big)h_{\mu\nu}=\frac{1}{2\ell^2}\big(\cD^L \cD^R h\big)_{\mu\nu}.
\end{equation}
For the remaining three operators, one can show the following identity
\begin{eqnarray}
\big(\cD^{p_1}\cD^{p_2}\cD^{p_3}h\big)_{\mu\nu}&=& h_{\mu\nu}
+\bigg(\frac{1}{p_1}+\frac{1}{p_2}+\frac{1}{p_3}\bigg)\bar{\epsilon}{_{\mu}}^{\alpha \beta}\bar{\nabla}_{\alpha} h_{\beta \nu}
+\frac{1}{p_1 p_2 p_3}\bar{\epsilon}{_{\mu}}^{\alpha \beta}\bar{\nabla}_{\alpha}\bigg( \bar{\square}+\frac{3}{\ell^2} \bigg)h_{\beta\nu}\nonumber \\
&&+\bigg( \frac{1}{p_1 p_2}+\frac{1}{p_1 p_3}+\frac{1}{p_2 p_3} \bigg)\big(\bar{\square}+\frac{3}{\ell^2}\big)h_{\mu\nu}.
\end{eqnarray}
Since all the operators commute, it is now easy to apply all of them to $h_{\mu\nu}$, which yields
\begin{eqnarray}
	\frac{1}{2 \ell^2}\big(\cD^L \cD^R \cD^{p_1}\cD^{p_2}\cD^{p_3}h\big)_{\mu\nu}&=&\cG_{\mu\nu}
	+\bigg(\frac{1}{p_1}+\frac{1}{p_2}+\frac{1}{p_3}\bigg)\bar{\epsilon}{_{\mu}}^{\alpha \beta}\bar{\nabla}_{\alpha} \cG_{\beta \nu}
	+\frac{1}{p_1 p_2 p_3}\bar{\epsilon}{_{\mu}}^{\alpha \beta}\bar{\nabla}_{\alpha}\bigg( \bar{\square}+\frac{3}{\ell^2} \bigg)\cG_{\beta\nu}\nonumber \\
	&+&\bigg( \frac{1}{p_1 p_2}+\frac{1}{p_1 p_3}+\frac{1}{p_2 p_3} \bigg)\big(\bar{\square}+\frac{3}{\ell^2}\big)\cG_{\mu\nu}.\label{eqD}
\end{eqnarray}
By inspection, one can see that the linearized equations (\ref{eqLin}) can be written in this form if the parameters ($p_1, p_2, p_3$) are chosen such that
\begin{eqnarray}
p_1+p_2+p_3&=&-\frac{a}{2 \mu},\nn \\
p_1 p_2+p_1 p_3+p_2 p_3&=&\frac{\bar{\sigma}}{\beta}-\frac{3}{\ell^2},\nn \\
p_1 p_2 p_3&=&\frac{2 \mu^2-\bar{\sigma}a}{2\beta\mu}+\frac{3a}{2\mu \ell^2}.
\end{eqnarray}
For generic values of the parameters, there is one set of real roots for ($p_1,p_2,p_3$), whose explicit expressions are complicated and not very illuminating to depict here as they solve a cubic equation. Since the operators defined in (\ref{operators}) commute, the most general solution for the equation (\ref{eqD}) can be written as
\begin{equation}
	h _ { \mu \nu } = h _ { \mu \nu } ^ { L } + h _ { \mu \nu } ^ { R } + h _ { \mu \nu } ^ { m _ { 1 } } + h _ { \mu \nu } ^ { m _ { 2 } }+ h _ { \mu \nu } ^ { m _ { 3 } },
\end{equation}
where
\begin{equation}
	( \mathcal { D }^L h^L ) _ { \mu \nu } = 0,\qquad	( \mathcal { D }^R h^R ) _ { \mu \nu } = 0,\qquad( \mathcal { D }^{p_i} h^{m_i} ) _ { \mu \nu } = 0,\qquad i = 1, 2, 3.
\end{equation}
Since $\big(\cD^L \cD^R h\big)_{\mu\nu}=0$ implies $\cG_{\mu\nu}=0$, $h _ { \mu \nu } ^ { L } $ and $h _ { \mu \nu } ^ { R } $ are the two massless excitations in the theory. But these are the modes that already exist in Einstein's theory, so they are pure gauge modes in the bulk. With the help of the following equation
\begin{equation}
	\left( \mathcal { D } ^ { - p } \mathcal { D } ^ { p } h \right) _ { \mu \nu } = - \frac { 1 } { p ^ { 2 } } \left( \bar { \square } + \frac { 3 } { \ell ^ { 2 } } - p ^ { 2 } \right) h _ { \mu \nu }, 
\end{equation}
it is easy to see that the remaining solutions describe massive excitations with the masses
\begin{equation}
	m _ { i } ^ { 2 } = p _ { i } ^ { 2 } - \frac { 1 } { \ell ^ { 2 } }.
\end{equation}
Since we have a real set of solutions for ($p_1, p_2, p_3$), the Breitenlohner-Freedman bound $m _ { i } ^ { 2 } \geq - \frac { 1 } { \ell ^ { 2 } }$ \cite{BF} is automatically satisfied and we have three nontachyonic massive excitations.

\section{Conserved Charges}

Having identified the spin-2 modes in the theory,  we now compute the energy and the angular momentum of the BTZ black hole by using the Abbott-Deser-Tekin technique \cite{ad,adt}. For a spacetime metric $g_{\mu\nu}$ having asymptotically  the same Killing symmetries as the background space, one can define ”conserved charges” from the linearized field equations which is of the following generic form
\begin{equation}
	\mathcal { O } ( \overline { g } ) _ { \mu \nu \alpha \beta } h ^ { \alpha \beta } = \kappa\, T _ { \mu \nu }.
\end{equation}
For each background Killing vector $\bar{ \xi}_\mu$, satisfying $\bar{\nabla}_{(\mu}\xi_{\nu)}=0$, a conserved current can be formed as
\begin{equation}
	\sqrt { - \overline { g } } \overline { \nabla } _ { \mu } \left( \overline { \xi } _ { \nu } T ^ { \mu \nu } \right) = \partial _ { \mu } \left( \sqrt { - \overline { g } } \overline { \xi } _ { \nu } T ^ { \mu \nu } \right) = 0.
\end{equation}
By applying Stokes' theorem, one obtains an expression for the conserved global charges
\begin{equation}
	Q ^ { \mu } ( \overline { \xi } ) = \int _ { \mathcal { M } } d ^ { n - 1 } x \sqrt { - \overline { g } } \overline { \xi } _ { \nu } T ^ { \mu \nu } = \int _ { \Sigma } d \Sigma _ { i } \mathcal { F } ^ { \mu i },
\end{equation}
where $\mathcal { M }$ is the ($n-1$)-dimensional spatial manifold, $\Sigma$ is its boundary and the antisymmetric tensor $\mathcal { F } ^ { \mu \nu }$ satisfies $T ^ { \mu \nu } \xi _ { \nu } = \overline { \nabla } _ { \nu } \mathcal { F } ^ { \mu \nu }$. Charge expressions for the $\cG$, $\epsilon \nabla \cG$ and $\square\cG$ terms in the linearized field equations (\ref{eqLin}) were obtained in \cite{ad}, \cite{dt_tmg} and \cite{adt} respectively. For the $\epsilon \nabla \square \cG $ term, one can make use of the equation
\begin{eqnarray}
2 \overline { \xi } ^ { \nu }\bar{\epsilon}{_{\mu}}^{\alpha\beta}\bar{\nabla}_{\alpha}\bar{\square}\cG_{\beta\nu} = \overline { \nabla } _ { \alpha } \left\{ \bar{\epsilon} ^ { \mu \alpha \beta } \overline { \square }\mathcal { G } _ { \nu \beta }  \overline { \xi } ^ { \nu } + \bar{\epsilon} ^ { \nu \alpha } _ { \beta } \overline { \square }\mathcal { G }  ^ { \mu \beta } \overline { \xi } _ { \nu } + \bar{\epsilon} ^ { \mu \nu \beta } \overline { \square }\mathcal { G } ^ { \alpha } _ { \beta } \overline { \xi } _ { \nu } \right\} + X_ { \beta } \overline { \square }\mathcal { G } ^ { \mu \beta },
\end{eqnarray}
and the final result can be written as
\begin{equation}
	Q ^ { \mu } ( \overline { \xi } ) = \frac { 1 } { 2 \pi G _ { 3 } } \oint _ { \partial \Sigma }\sqrt { - \bar  { g } }\,  d l _ { i }\, q ^ { \mu i }( \overline { \xi } ), 
\end{equation}
where
\begin{eqnarray}
q^{\mu i}( \overline { \xi } )&=&\left(1-\frac{\bar{\sigma}a}{2\mu^2}\right) q^{\mu i}_{(1)}( \overline { \xi } )+\frac{\bar{\sigma}}{2\mu}\left[q^{\mu i}_{(1)}( \overline { X } )+q^{\mu i}_{(2)}( \overline { \xi } )\right]+\nn \\
&&-\frac{\beta a}{2 \mu^2}q^{\mu i}_{(3)}( \overline { \xi } )+\frac{\beta}{2\mu}\left[q^{\mu i}_{(3)}( \overline { X } )+q^{\mu i}_{(4)}( \overline { \xi } )\right].\nn\\
q^{\mu i}_{(1)}( \overline { \xi } ) &=&\overline { \xi } _ { \nu } \overline { \nabla } ^ { \mu } h ^ { i \nu } - \overline { \xi } _ { \nu } \overline { \nabla } ^ { i } h ^ { \mu \nu } + \overline { \xi } ^ { \mu } \overline { \nabla } ^ { i } h - \overline { \xi } ^ { i } \overline { \nabla } ^ { \mu } h  \nn \\
&&+ h ^ { \mu \nu } \overline { \nabla } ^ { i } \overline { \xi } _ { \nu } - h ^ { i \nu } \overline { \nabla } ^ { \mu } \overline { \xi } _ { \nu } + \overline { \xi } ^ { i } \overline { \nabla } _ { \nu } h ^ { \mu \nu } - \overline { \xi } ^ { \mu } \overline { \nabla } _ { \nu } h ^ { i \nu } + h \overline { \nabla } ^ { \mu } \overline { \xi } ^ { i }, \nn \\
q^{\mu i}_{(2)}( \overline { \xi } )&=&\bar{\epsilon} ^ { \mu i \beta } \mathcal { G } _ { \nu \beta } \overline { \xi } ^ { \nu } + \bar{\epsilon} ^ { \nu i \beta } \mathcal { G } _ { \beta } ^ { \mu } \overline { \xi } _ { \nu } + \bar{\epsilon} ^ { \mu \nu \beta } \mathcal { G } ^ { i } _ { \beta } \overline { \xi } _ { \nu }, \nn \\
q^{\mu i}_{(3)}( \overline { \xi } )&=&\overline { \xi } _ { \nu } \overline { \nabla } ^ { i } \mathcal { G }  ^ { \mu \nu } - \overline { \xi } _ { \nu } \overline { \nabla } ^ { \mu } \mathcal { G }  ^ { i \nu } - \mathcal { G } ^ { \mu \nu } \overline { \nabla } ^ { i } \overline { \xi } _ { \nu } + \mathcal { G }  ^ { i \nu } \overline { \nabla } ^ { \mu } \overline { \xi } _ { \nu },\nn\\
q^{\mu i}_{(4)}( \overline { \xi } )&=&\bar{\epsilon} ^ { \mu i \beta } \overline { \square }\mathcal { G } _ { \nu \beta } \overline { \xi } ^ { \nu } + \bar{\epsilon} ^ { \nu i \beta } \overline { \square }\mathcal { G } _ { \beta } ^ { \mu } \overline { \xi } _ { \nu } + \bar{\epsilon} ^ { \mu \nu \beta } \overline { \square }\mathcal { G } ^ { i } _ { \beta } \overline { \xi } _ { \nu },\label{charge}
\end{eqnarray}
and $\bar{X} ^ { \beta } = \epsilon ^ { \alpha \nu \beta } \overline { \nabla } _ { \alpha } \overline { \xi } _ { \nu }$ is also a background Killing vector.

Let us now apply the above construction to find the charges of the the rotating BTZ black hole in this theory. BTZ is locally AdS$_3$ and hence it is a solution of the theory once the cosmological constant is adjusted.  In the usual $(t,r,\phi)$ coordinates, the metric reads 
\begin{equation}
	d s ^ { 2 } = \left( m G _ { 3 } + \Lambda r ^ { 2 } \right) d t ^ { 2 } - j d t d \phi + r ^ { 2 } d \phi ^ { 2 } + \frac { d r ^ { 2 } } { - m G _ { 3 } - \Lambda r ^ { 2 } + \frac { j ^ { 2 } } { 4 r ^ { 2 } } },
\end{equation}
where the background metric is found by setting $m = 0$ and $j = 0$ as
\begin{equation}
d s ^ { 2 } = \Lambda r ^ { 2 } d t ^ { 2 } + r ^ { 2 } d \phi ^ { 2 } - \frac { d r ^ { 2 } } { \Lambda r ^ { 2 } }.
\end{equation}
In the asymptotic region, the linearized cosmological Einstein tensor vanishes $\cG_{\mu\nu}=0$ and only $q^{\mu i}_{(1)}$ terms in (\ref{charge}) contribute. Killing vectors $\overline { \xi } ^ { \mu } = - \left( \frac { \partial } { \partial t } \right) ^ { \mu }$   and  $\overline { \xi } ^ { \mu } = \left( \frac { \partial } { \partial \phi } \right) ^ { \mu }$ yield the energy and the angular momentum, respectively, as
\begin{equation}
	E = \frac { 1 } { G _ { 3 } } \left[\left(1-\frac{\bar{\sigma}a}{2 \mu^2}\right) m + \frac { j \Lambda \bar{\sigma} } { \mu } \right] , \quad J = \frac { 1 } { G _ { 3 } } \left[ \left(1-\frac{\bar{\sigma}a}{2\mu^2}\right) j - \frac { m\bar{\sigma} } { \mu } \right]
\end{equation}

\section{CONCLUSIONS}
In three dimensions, given a symmetric 2-tensor, say $L_{\mu\nu}$, one can construct another symmetric 2-tensor by taking the ``curl" of the former as
\begin{equation}
H_{\mu\nu} := \epsilon_{\mu}{^{\alpha\beta}}\nabla_{ \alpha }L_{\beta\nu}+\epsilon_{\nu}{^{\alpha\beta}}\nabla_{\alpha}L_{\beta\mu}
\end{equation}
when $L_{\mu\nu}$ is the Einstein tensor ($G_{\mu\nu}$), the $H_{\mu\nu}$ tensor becomes the Cotton-York tensor ($C_{\mu\nu}$) which is traceless, divergence-free. The latter fact yields the topologically massive gravity $\big(G_{\mu\nu}+\frac{1}{\mu}C_{\mu\nu}=0\big)$ as a consistent theory. But when $L_{\mu\nu}$ is taken as the Cotton-York tensor $C_{\mu\nu}$, the resulting $H_{\mu\nu}$ tensor, even though it is traceless, it is not divergence-free. So the curl of the Einstein tensor (which is the Bach tensor in 3D) is not a conserved tensor. But here we have given a full construction of how one can start from a divergence-free, symmetric tensor ($\Psi_{\mu\nu}$) and write an on-shell covariant theory by taking the curl of $\Psi_{\mu\nu}$ and by adding (judiciously chosen) quadratic terms in $\Psi_{\mu\nu}$. This Bachian gravity is highly constrained and it should always involve Einstein's theory at the lowest order: hence it is a deformation of the 2+1-dimensional general relativity.
We have given examples of $\Psi_{\mu\nu}$ coming from the quadratic gravity, carried out the linearized field equations and computed the particle content of the theory, as well as conserved charges of the BTZ black hole. We also explained how $f(\text{Ricci})$-type extensions can be found by giving a cubic theory as an example. A Born-Infeld-type extension of the quadratic actions that yield consistent field equations can also be considered. Indeed, one of the combinations ($4 \alpha + \frac { 3 } { 2 } \beta = 0$) that we found in (\ref{sol}) defines the NMG theory and BINMG theory described by the action
\begin{equation}
	I _ { \mathrm { BINMG } } = - \frac { 4 m ^ { 2 } } { \kappa ^ { 2 } } \int d ^ { 3 } x \left[ \sqrt { - \operatorname { det } \left( g + \frac { \sigma } { m ^ { 2 } } G \right) } - \left( 1 - \frac { \lambda _ { 0 } } { 2 } \right) \sqrt { - \operatorname { det } g } \right],
\end{equation}
was shown to give the same combination with redefined parameters at the linearized level. For the second combination in (\ref{sol}), finding a Born-Infeld-type extension is still an open problem.  A naive extension of these ideas, that is constructing on-shell consistent non-trivial theories say with massive gravitons, in four dimensions is not immediate: the curl of a symmetric 2-tensor is not a 2-tensor but a 3-tensor.

\end{document}